\documentclass[11pt,twocolumn,superscriptaddress,longbibliography,reprint]{revtex4-1}

\usepackage[mathlines]{lineno}
\usepackage{cancel,color}
\usepackage{bm}

\usepackage{wrapfig}
\usepackage{graphicx}
\graphicspath{{pics/}}

\usepackage{soul} 

\usepackage{url}

\newcommand{\be}{\begin{eqnarray}}
\newcommand{\ee}{\end{eqnarray}}

\newcommand{\fig}[6]
{
\begin{#1}[#2]
\centering
    \includegraphics[width=#3\linewidth]{#4}\caption{#5}\label{#6}
\end{#1}
} 
\begin{document}

\title{Inverse Design of Broadband Antennas for Terahertz Devices Based on 2D Materials}


\author{M. Y. Lukianov}
\affiliation{Programmable Functional Materials Lab, Center for Neurophysics and Neuromorphic Technologies, Moscow, 127495}

\author{A. Maevskiy}
\affiliation{Institute for Functional Intelligent Materials, National University of Singapore, Singapore, 117544, Singapore}

\author{N. Kazeev}
\affiliation{Institute for Functional Intelligent Materials, National University of Singapore, Singapore, 117544, Singapore}

\author{D. Mylnikov}
\affiliation{Moscow Center for Advanced Studies, Kulakova str. 20, Moscow, 123592, Russia}

\author{D.A.~Svintsov}
\affiliation{Moscow Center for Advanced Studies, Kulakova str. 20, Moscow, 123592, Russia}

\author{K.S. Novoselov}
\affiliation{Institute for Functional Intelligent Materials, National University of Singapore, Singapore, 117544, Singapore}


\author{A. Ustyuzhanin}
\affiliation{Constructor University, Bremen, Campus Ring 1, 28759, Germany}
\affiliation{Institute for Functional Intelligent Materials, National University of Singapore, Singapore, 117544, Singapore}

\author{D.A. Bandurin$^{*}$}
\affiliation{Institute for Functional Intelligent
Materials, National University of Singapore, Singapore, 117544, Singapore}
\affiliation{Programmable Functional Materials Lab, Center for Neurophysics and Neuromorphic Technologies, Moscow, 127495}
\affiliation{Department of Materials Science and Engineering, National University of Singapore, 117575 Singapore}
\email{Correspondence to: dab@nus.edu.sg}


\begin{abstract}
Terahertz (THz) technology, a cornerstone of next-generation high-speed communication and sensing, has long been hindered by impedance mismatch challenges that limit device performance and applicability. These challenges become particularly pronounced when ultrasensitive two-dimensional (2D) materials are employed as the device substrate in the THz range, further complicating their integration into real-world applications. Furthermore, conventional antenna designs often fail to provide adequate matching across the extensive THz spectrum. In this work, we tackle these challenges using a procedural generation algorithm to design THz broadband antennas that satisfy specific performance criteria. Namely, the developed inverse design methodology enables customization for the target impedance value, bandwidth, and contact topology requirements. The proposed antenna achieves an improvement of up to 40\% in power transfer efficiency compared to traditional bow-tie antennas under realistic operating conditions. High-fidelity electromagnetic simulations validate these results, confirming the design's practicality for THz applications. This work addresses critical limitations of existing antenna designs and advances the feasibility of high-frequency applications in both communication and sensing.
\end{abstract}

\maketitle

Transmitting, receiving, and coupling electromagnetic (EM) signals in the terahertz (THz) range pose a fundamental challenge for advancing high-speed communication and sensing technologies. This frequency region, spanning from 0.1 to 3 THz, and located between microwaves and infrared light, has spurred extensive research efforts~\cite{bridgeDAgap}.
As data traffic and bandwidth demands grow, reaching carrier frequencies beyond 100 GHz is increasingly necessary to support data rates above 100 Gbps~\cite{comGo100Thzfirst,comGo100Thz2,comGo100Gbs}. Recent innovations in electronic and photonic THz transceivers have enabled more efficient signal generation, modulation, and directivity, facilitating practical advancements in both communication and sensing applications~\cite{transvr1,transvr2,transvr3,transvr4,transvr5}.

The versatility of THz technology spans communication, imaging, and sensing, blurring the boundaries between research and commercial applications, each demanding trade-offs in sensitivity, speed, resolution, and bandwidth~\cite{2021imaging,imag2comm}.
Across these applications, the role of efficient interfaces is crucial, as the impedance mismatch between antennas and devices can severely hinder performance~\cite{joint2024terahertz}.
Conventional RF antennas are typically designed for a 50~$\Omega$ impedance~\cite{huangbook}, while new generation of promissing THz detectors based on van der Waals materials \cite{koppens2014photodetectors, mittendorff20212d} often exhibit much higher impedance values (see Figure~\ref{fig:pipeline_abstract}a for schematic).
For instance, detectors based on graphene \cite{vicarelli2012graphene, bandurin2018dual, 2018resonant, gayduchenko2021tunnel, castilla2019fast, kravtsov2024anomalous, soundarapandian2024high, koppens2014photodetectors, spirito2014high, ludwig2024terahertz} or black phosphorus~\cite{viti2016efficient, viti2015black} can have impedance values exceeding several k$\Omega$ in monolayer graphene, while bilayer graphene devices, although featuring exceptional THz responsivity~\cite{titova2023ultralow, mylnikov_limiting_2022}, may reach 100-500~$\Omega$. 
While these material platforms are advantageous for highly sensitive THz detectors, such high impedance values introduce significant design challenges.
Despite the recognition of the impedance mismatch problem since the late 1970s~\cite{hwang1979planar}, most THz antennas continue to adapt scaled-down microwave designs, such as dipole and bow-tie antennas. These designs typically have low impedance, leading to severe mismatches with high-impedance THz devices~\cite{mainreview}, limiting their performance across diverse applications.

Antenna shape optimization has long been used to improve radiation properties, but it typically fails to address the impedance mismatch directly. This limitation is often managed by impedance matching networks, which, for THz frequencies, require custom-designed circuit-integrated components. However, creating these networks for THz frequencies, especially across a wide frequency band (e.g., 0.2–1.2 THz), is exceptionally complex. Although research has demonstrated the possibility of complete impedance matching for two-dimensional materials at specific wavelengths~\cite{mylnikov_limiting_2022}, these solutions are limited to single frequencies, restricting their applicability for wideband systems. A more feasible strategy is to design custom antenna shapes through lithography, which offers a less complex solution than fabricating THz-specific impedance matching networks.

\fig{figure*}{htb}{1}{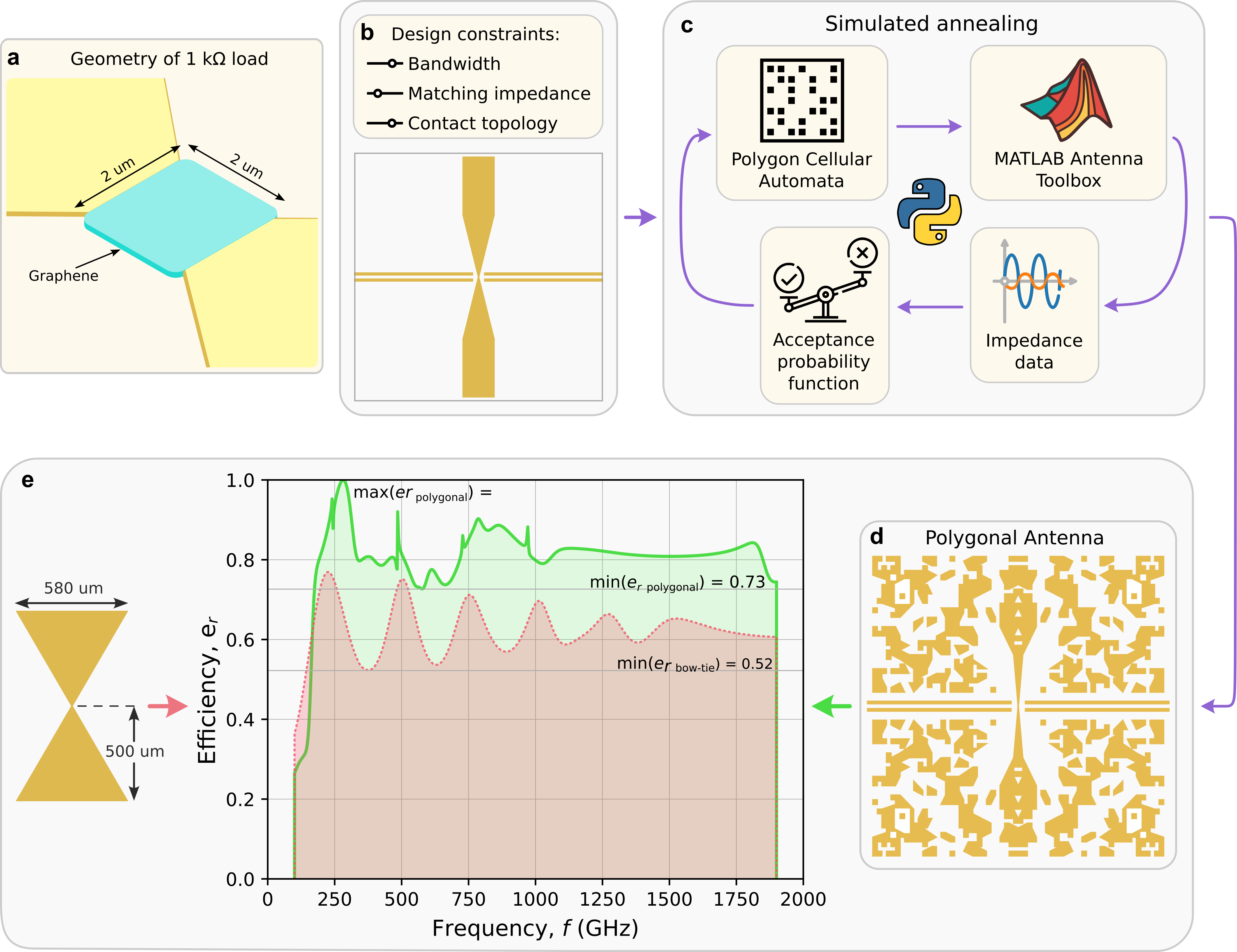}
{\textbf{Electrodynamic modeling of the antenna efficiency profiles.}
\textbf{a,} Geometry of 1~$\mathrm{k}\Omega$ load used in place of a device.
\textbf{b,} Input design constraints that guide the optimization process.
\textbf{c,} Method workflow detailing the step-by-step procedure.
\textbf{d,} Resulting shape of the optimized polygonal antenna.
\textbf{e,} Efficiency as a function of frequency, as simulated in CST Studio Suite for both bow-tie antenna (red line) and polygonal antenna (green line)}{fig:pipeline_abstract}

The generation of such antennas often relies on advanced optimization techniques~\cite{2023R_sarkar_CIforUWB}, including genetic algorithms~\cite{holland1992genetic}, particle swarm optimization~\cite{kennedy1995particle}, simulated annealing (SA)~\cite{kirkpatrick1983sa}, and neural networks~\cite{bishop1994neural}. These methods provide powerful tools for navigating vast and intricate design spaces to achieve superior performance~\cite{vasylenko2022natural}.
Among these approaches, Convolutional Neural Networks (CNNs) have demonstrated remarkable efficacy in inverse design tasks within nanophotonics, enabling the identification of structures tailored to precise optical properties~\cite{molesky2018inverse}. Despite this progress, the application of such advanced machine learning methodologies to the design and optimization of 2D-material-based terahertz devices remains largely underexplored, presenting a compelling opportunity for innovation in this domain.

In this work, we propose a method for generating custom polygonal WB antennas for THz applications, guided by SA to minimize impedance mismatch while satisfying bandwidth and contact topology constraints
(Figure~\ref{fig:pipeline_abstract}b).
Our algorithm begins with an initial antenna shape that is refined iteratively, using simulation to evaluate performance at each step (Figure 1c). This design synthesis approach allows for efficient impedance matching across a broad frequency range, mitigating limitations found in conventional antenna designs. Finally, the effectiveness of our optimized designs is confirmed through high-fidelity electromagnetic simulations, demonstrating a practical solution to the long-standing impedance mismatch issue in THz technology.

\section*{Results}
We first outline the pipeline of our antenna generation and optimization algorithm, providing a detailed account of each step. The process begins with a polygonal cellular automata algorithm that modifies an initial antenna shape. In each iteration cycle, the shape undergoes a single stochastic transformation, allowing the algorithm to systematically explore the design space. The modified design is then subjected to electrodynamic finite element simulation to evaluate its performance. The simulation results are processed using an acceptance probability function, which determines whether the design is retained or discarded, thereby concluding the iteration. SA is integrated into this framework to guide the exploration process and ensure convergence toward optimal solutions. Notably, the optimized antenna demonstrates a 40\% improvement in power efficiency (from 0.52 to 0.73), underscoring SA as an effective optimization strategy in this context. A comprehensive description of the SA methodology is provided in the subsequent section.
\newline

\textbf{Simulated Annealing.} SA, an optimization technique inspired by the physical process of annealing in metallurgy, is a cornerstone of our antenna design pipeline. In metallurgy, annealing involves heating a material to a high temperature and gradually cooling it to remove defects and stabilize its structure. Similarly, in optimization, SA simulates this cooling process in attempt to balance exploration of the solution space with the refinement of promising configurations.

The SA process begins with the evaluation of an initial antenna shape that combines features of dipole and bow-tie antennas, enabling diverse evolution paths. In each iteration, the algorithm generates a new candidate shape by introducing stochastic modifications to the current design. These transformations alter the geometry by expanding or contracting the structure and mixing pixelated clusters with 2D noise. The candidate shape is then re-evaluated to determine its performance.

Candidate acceptance is governed by the loss function, as defined later in the text, and the Boltzmann probability function (Equation~\ref{eq:probability}). If the candidate achieves a lower loss value, it replaces the current design as the optimal solution. If the loss increases, the candidate may still be accepted with a probability depending on the temperature. This probabilistic acceptance mechanism mitigates the risk of the algorithm becoming trapped in local minima. The probability function is defined as follows:

\begin{equation}
    \label{eq:probability}
    P = \exp\left(-\frac{\Delta \text{Loss}^+}{T}\right),
    \hspace{5pt}\Delta\text{Loss}^+\equiv\max\left(\Delta\text{Loss},0\right)
\end{equation}

The temperature is reduced according to an exponential decay schedule, gradually transitioning the algorithm's focus from exploration to exploitation. This cooling schedule prioritizes fine-tuning in the vicinity of the most promising solutions as optimization progresses. The process continues until a predefined stopping criterion is met, such as reaching a maximum number of iterations, a temperature threshold, or stagnation in performance improvements. At the conclusion of the optimization, the antenna shape with the lowest loss value encountered during the process is selected as the final design.
\newline

\subsection{Simulation Details}

\textbf{Simulation Details.} The evaluation of antenna designs within the optimization loop was conducted using MATLAB Antenna Toolbox. It was chosen for its compatibility with several critical requirements, including a fully supported Python interface, a frequency-domain solver for radiation and scattering problems, support for custom geometries, and a feed model with customizable impedance.

Simulation parameters were configured to reflect realistic device specifications, targeting specific impedance, bandwidth, and frequency ranges. The feeding cross-section was defined as 1~{\textmu}m in length, matched to an impedance of 1~$\mathrm{k}\Omega$. Gold metallic components were modeled as Perfect Electric Conductors, simplifying computational complexity while maintaining physical accuracy. To balance computational efficiency and precision, meshing parameters were set with a minimum edge length equal to one-eighth of the smallest wavelength. Although coarser than MATLAB’s one-tenth wavelength guideline~\cite{mlabMeshing}, this configuration provided a practical compromise between resource demands and the accuracy required to estimate antenna designs at each iteration.

A comparative analysis between MATLAB Antenna Toolbox and CST Studio simulations revealed notable differences in the efficiency profiles (Figure~\ref{fig:antenna_evo} (c) and (e)). CST Studio simulations offered higher fidelity and more precise results but required significantly longer computation times. Conversely, MATLAB simulations prioritized faster processing, albeit with reduced accuracy. This trade-off underscores the role of MATLAB in expediting iterative optimization cycles, while CST Studio functioned as the benchmark for high-fidelity validation of final antenna designs.
\newline


\textbf{Loss Function.} The loss function (Eq.~\ref{eq:loss}) evaluates antenna performance by mapping complex impedance values into a scalar metric, quantifying the antenna's ability to absorb electromagnetic energy over a specified frequency range. Optimal power transfer is achieved when the device impedance, \(Z_d = R_d + i X_d\), matches the complex conjugate of the antenna impedance, \(Z_a^*\), a principle known as conjugate matching~\cite{balanis, milligan}. The power transfer efficiency, \(e_r = 1 - |\Gamma|^2\), depends on the reflection coefficient, \(\Gamma\), defined as:

\begin{equation}
    \label{eq:gamma}
    \Gamma= \frac{Z_a-Z_d}{Z_a+Z_d}
\end{equation}
For our analysis, we used a purely resistive device load of 1~$\mathrm{k}\Omega$ as the benchmark, with the device impedance set to \(R_d = 1~\mathrm{k}\Omega\) and \(X_d = 0\). Under these conditions, maximum power transfer occurs when \(\Gamma\) approaches zero.

To ensure robust performance across the frequency range, the loss function emphasizes maximizing $e_r$ while mitigating the impact of sharp maxima in reflection coefficients. The antenna impedance is sampled over $N$ linearly spaced frequencies. The corresponding reflection coefficients $\Gamma_i$ are sorted in decreasing order, and $K$ largest values contribute to the loss function as follows:

\begin{equation}
    \label{eq:loss}
    \text{Loss} = \frac{1}{K}\sum_i^K (0.3 + \Gamma_i)^{3}
\end{equation}
The offset of 0.3 and the cubic weighting scale the loss values to a practical range of [0.027, 2.197], centering the benchmark bow-tie antenna's reflection coefficient of 0.7 at a loss value of 1.0.

We adopted the setting of $K = N/10$, which results in a smoother optimization process. In contrast, setting $K = 1$, i.e., minimizing the largest reflection coefficient, led to fluctuations in loss values caused by minor geometric changes, disrupting the shape evolution.
When $K = N$, the approach is equivalent to averaging across all reflection coefficients, but this method is uninformative as it fails to emphasize critical maxima or oscillating regions in the frequency spectrum.
\newline

\textbf{Shape representation and modification operations.} The antenna shape is represented as a binary 32x32 matrix (Figure~\ref{fig:gen_matrix}), which allows for stochastic transformations during the optimization process. The matrix corresponds to a quarter of the full antenna obtained by applying horizontal and vertical reflection operations (Figure~\ref{fig:gen_polygons}, step 11).

\fig{figure}{ht!}{1}{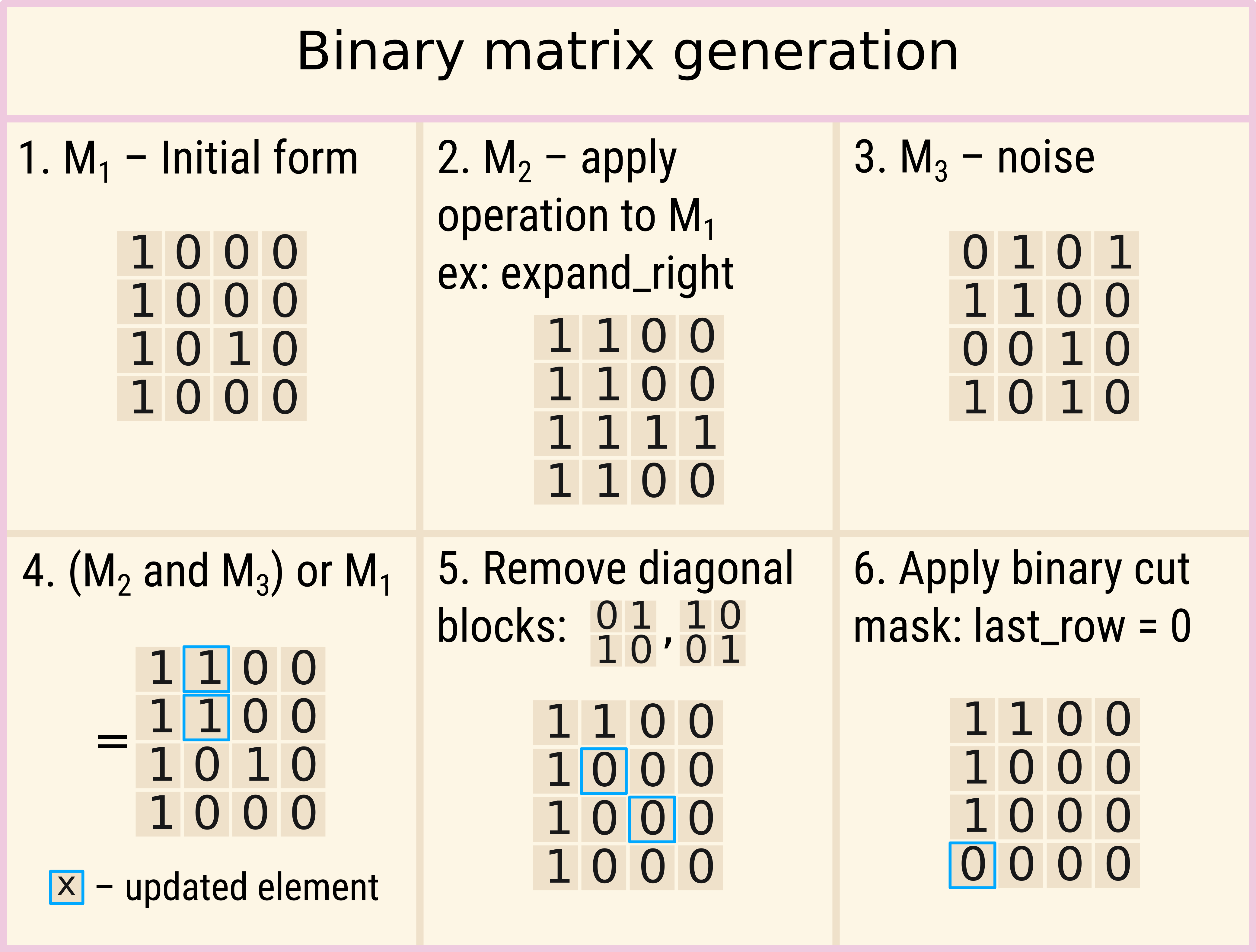}{\textbf{Binary Matrix Procedural Generation Process.} The algorithm consists of six sequential steps, each representing a specific transformation to generate diverse patterns in a controlled manner. These transformations enable the creation of new shapes from existing ones. For illustration, a simplified 4x4 matrix example is shown, while a 32x32 grid was used during actual optimization.}{fig:gen_matrix}

\fig{figure}{hb!}{0.9}{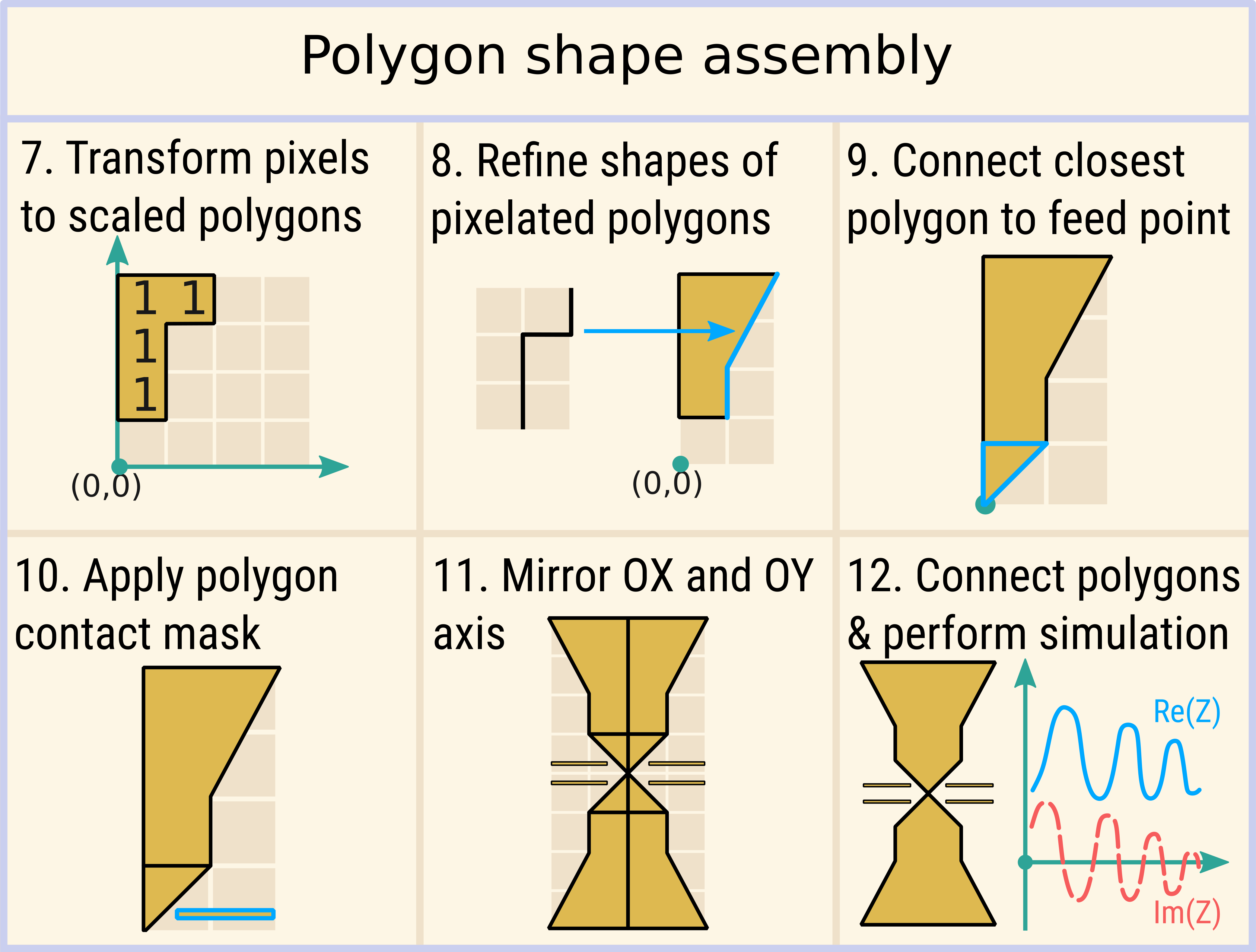}{\textbf{Polygon shape assembly: from binary matrix to EM modeling.} The figure illustrates a six-step process that transforms a binary matrix into a finalized polygonal antenna shape. Steps include: (7-9) Transformation of the binary matrix into a set of polygons; (10) Addition of adjacent contacts in the vicinity of the load; (11) Application of symmetry to the polygonal structure. (12) EM simulation of the resulting shape, yielding the complex impedance as a function of frequency.}{fig:gen_polygons}

\fig{figure*}{ht}{1}{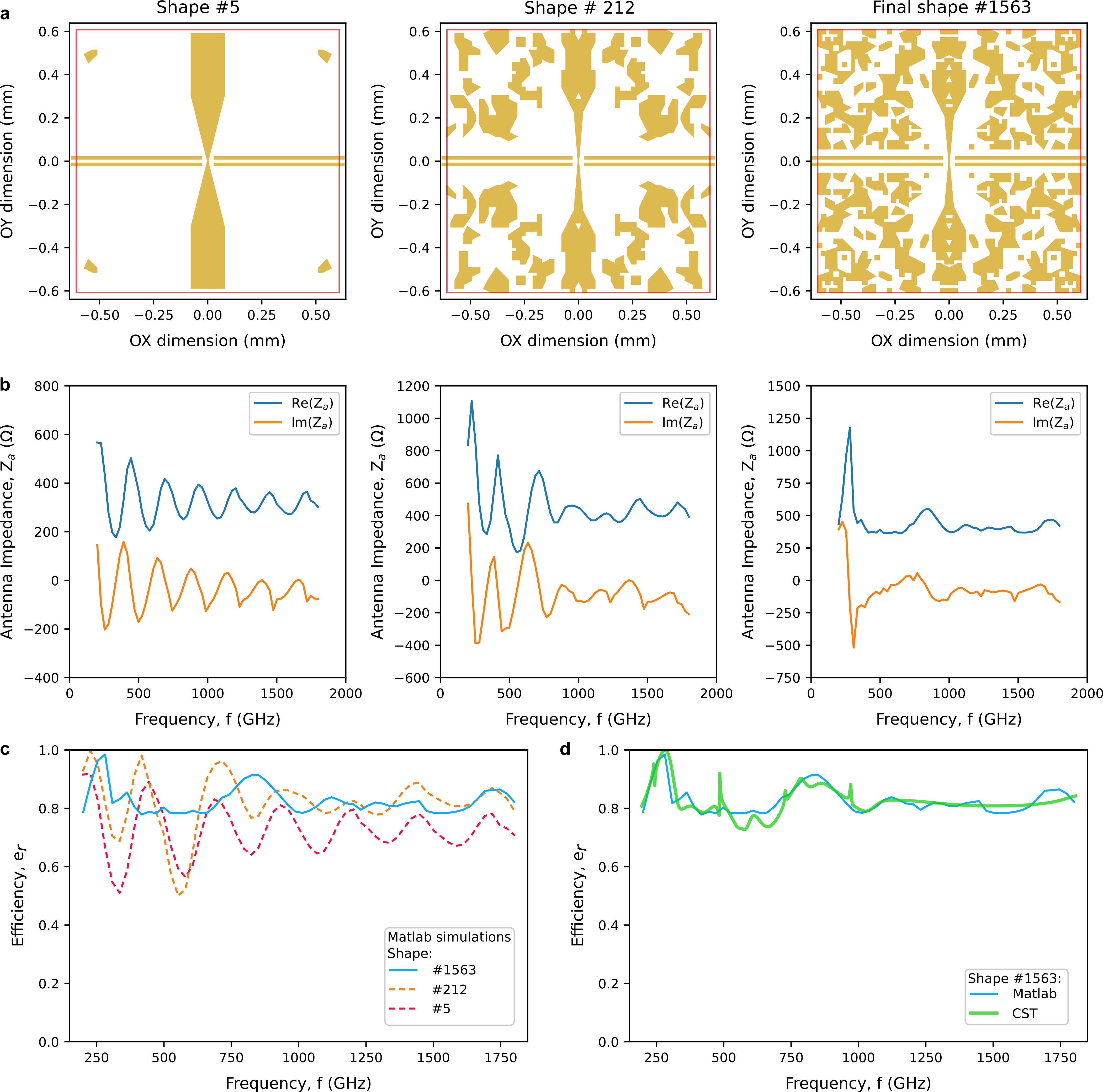}{\textbf{Electrodynamic modeling and evolution of polygonal antennas through simulated annealing optimization.}  \textbf{a,} From left to right: Three stages of antenna evolution, with the final design shown in the last generation. \textbf{b,} Complex impedance spectra corresponding to the three designs. \textbf{c,} Efficiency profiles as a function of frequency simulated in MATLAB. The dashed red line represents shape $\#5$, the dashed orange line represents shape $\#212$, and the solid blue line corresponds to the final shape. \textbf{d,} Comparison of efficiency profiles for shape $\#1563$ simulated using MATLAB (blue line) and CST studio suite (green line).}{fig:antenna_evo}

Shape evolution begins with stochastic transformations applied to a pixel-based binary matrix, as illustrated in Figure~\ref{fig:gen_matrix}. This step uses a cellular automata-inspired algorithm to iteratively modify the pixelated structure through expanding or shrinking the shape by a single pixel in a random direction and adding or removing isolated pixel clusters (Figure~\ref{fig:gen_matrix}, step 2).
To introduce further variability, fractal noise is integrated into this process, enabling the creation of intricate concave and perforated structures with minimal modification steps (Figure~\ref{fig:gen_matrix}, step 3). Together, these methods provide sufficient variability to explore a wide design space and generate diverse geometries.

Once the pixel transformations are complete, the binary matrix is converted into polygons, as shown in Figure~\ref{fig:gen_polygons}. The transformation process begins by labeling each region of the binary matrix, distinguishing between solid structures and holes (Figure~\ref{fig:gen_polygons}, step 7). Next, the labeled polygons undergo a contour refinement process to smooth sharp transitions and minimize abrupt periodicity (Figure~\ref{fig:gen_polygons}, step 8). This refinement improves the antenna performance by mitigating localized impedance mismatches, resulting in more uniform efficiency across the frequency spectrum.

To further enhance antenna performance, special attention is given to the feeding connection within the polygonal structure, a region of high current density. Bow-tie-like feeding configurations are shown to provide superior bandwidth compared to fixed rectangular feeds. A specialized function was developed to optimize feed placement by locating the nearest polygon, identifying the closest and most rightward points, and integrating these coordinates into the feed design (Figure~\ref{fig:gen_polygons}, step 9).

The finalized polygons are merged using binary operations and sent to the simulation engine for electrodynamic analysis, completing the transformation workflow (Figure~\ref{fig:gen_polygons}, step 12). The electromagnetic simulation provided complex impedance as a function of frequency, which is used to obtain the loss value, as discussed above.

\section*{Discussion}

To evaluate the performance of the proposed designs, a classic bow-tie antenna was selected as a benchmark due to its established versatility in wideband sensing~\cite{2018resonant} and communication~\cite{uwbGaAs2017} applications. As shown in figure~\ref{fig:pipeline_abstract}e, the polygonal design achieves, on average, 40\% higher efficiency compared to the bow-tie antenna. Additionally, the polygonal antenna exhibits a near-constant real impedance of approximately 400~$\Omega$, a critical characteristic for sensing applications.

The observed impedance, while lower than the target of 1~$\mathrm{k}\Omega$, reflects inherent trade-offs in the design and optimization process. Balancing multiple objectives, such as efficiency, bandwidth, and shape diversity, occasionally overrides strict adherence to the impedance target. The optimization framework prioritizes a flatter efficiency profile, which may favor overall performance rather than exact impedance matching. Furthermore, the physical design area, constrained to a 1,215 by 1,215 micrometer square, inherently limits the achievable capacitance, resistance, and inductance, thereby bounding the range of possible matching impedances.

The optimization progression is illustrated in Figure~\ref{fig:antenna_evo}, which depicts three polygonal antennas at different steps, including their shapes, impedance profiles, and efficiency across frequencies. The results reveal improved efficiency profiles with reduced oscillations and a higher mean value as the optimization advances. Notably, at 283 GHz, a sharp efficiency peak indicates near-perfect coupling between the antenna and the device, corresponding to the reflection coefficient \(\Gamma\) reaching its minimum when \(Z_\mathrm{a}\) matches \(Z_\mathrm{d}\).
This phenomenon persisted across iterations, even though the loss function prioritized smoothing sharp minima in efficiency profiles and disregarded sharp maxima. These findings highlight an instance of perfect impedance matching. They  suggest that for applications requiring specific resonant efficiency peaks, the loss function could be adapted to explicitly prioritize such features.

Our proposed framework is adaptable to diverse device requirements, offering a modular approach to antenna design. By leveraging external simulation engines, the framework supports integration with various numerical techniques. Simulated annealing plays a pivotal role, offering flexibility by optimizing antennas without requiring differentiable or continuous loss functions, making it particularly effective for randomly generated antenna geometries.

Practical considerations also reinforce the feasibility of polygonal designs. The geometries were specifically optimized for reliable prototyping using standard lithographic techniques. We proactively avoided issues, such as polygons connected only through single vertices (which could cause unintended connections during fabrication), ensuring a smooth transition from simulation to physical realization.

\section*{Conclusion}

Our work introduces an innovative approach to THz antenna design, addressing critical impedance mismatch challenges, particularly for 2D-materials-based terahertz devices. The proposed wideband (WB) antenna, optimized for a frequency range of 0.2 to 1.8 THz, outperforms conventional designs such as the bow-tie antenna. The simulation results demonstrate an improvement in power efficiency 40\% and a near-constant real impedance of approximately 400~$\Omega$. In contrast, the bow-tie design exhibits significant impedance fluctuations, with an average value of 245~$\Omega$.

Beyond its demonstrated performance improvements, the polygonal design methodology offers a versatile framework adaptable to a variety of high-frequency device requirements. By integrating advanced optimization techniques with scalable simulation workflows, this approach lays the groundwork for future innovations in multi-functional antenna systems, such as those required for hybrid sensing-communication platforms or dynamic, reconfigurable THz networks.

This work underscores the potential of advanced shape optimization techniques for THz applications. The demonstrated improvements in power efficiency and impedance stability across a broad frequency range highlight the effectiveness of the proposed approach. Future efforts may focus on on-substrate integration, reduction of computational overhead and exploration of alternative topologies to further enhance the capabilities of THz sensing and communication systems.

Future work should prioritize integrating substrate effects into the optimization process to ensure that substrate geometry and material properties are systematically accounted for throughout the design pipeline. This approach would significantly enhance the framework's comprehensiveness, potentially boosting antenna performance and expanding its applicability to real-world scenarios. However, implementing such substrate-aware optimization poses challenges due to the substantial computational power required, which remains a limitation with current resources. Future efforts could also focus on strategies to minimize computational overhead and explore alternative topologies to further advance the capabilities of THz sensing and communication systems.

\section*{Acknowledgments}

The research at NUS is supported by MOE Tier 2 (Award T2EP50123-0020) and  A*STAR under its Young Individual Research Grant (Award M22K3c0106. M.Y.L. acknowledges the support of an internal funding program from the Center for Neurophysics and Neuromorphic Technologies. K.S.N. is grateful to the Ministry of Education, Singapore (Research Centre of Excellence award to the Institute for Functional Intelligent Materials, I-FIM, project No. EDUNC-33-18-279-V12) and to the Royal Society (UK, grant number RSRP\textbackslash R\textbackslash190000) for support. N.K., and A.U. acknowledge the support from the National Research Foundation Singapore under its AI Singapore Programme (Award Number: AISG3-RP-2022-028). D.A.M. acknowledges the support of the Russian Science Foundation, grant \#24-79-10081.

M.Y.L., A.M., N.K., K.S.N., A.U., and D.A.B. acknowledge the support from the Ministry of Education, Singapore, under its Research Centre of Excellence award to the Institute for Functional Intelligent Materials, National University of Singapore (I-FIM, project No. EDUNC-33-18-279-V12)

\bibliographystyle{ieeetr}
\bibliography{lit.bib}
\end{document}